\documentclass[journal,comsoc]{IEEEtran}

\usepackage{cite}

\usepackage{amsmath,amssymb,amsfonts}
\usepackage{algorithmic}
\usepackage{graphicx}
\usepackage{textcomp}
\usepackage{xcolor}
\usepackage{float}
\usepackage[nolist]{acronym}
\usepackage[none]{hyphenat}
\let\labelindent\relax
\usepackage[shortlabels]{enumitem}
\usepackage{tabularx}
\usepackage{comment}
\usepackage[normalem]{ulem}
\usepackage{url}

\usepackage{cancel}

\usepackage{placeins}
\usepackage{multirow}
\usepackage{longtable}
\usepackage{makecell}

\def\BibTeX{{\rm B\kern-.05em{\sc i\kern-.025em b}\kern-.08em
    T\kern-.1667em\lower.7ex\hbox{E}\kern-.125emX}}

\newif\ifrev
\revfalse 
\ifrev

\newcommand{\ricrev}[1]{\textcolor{blue}{#1}}

\else
    
\newcommand{\ricrev}[1]{#1}

\fi

\begin{document}

\title{Analysis of Measurements in a Microwave Oven with High Level of Non-Ionizing Radiation (NIR) Before and After Repair}

\author{Júlia~da~L.~A.~Silva,~Vicente~A.~de~Sousa~Jr.,~Marcio~E.~C.~Rodrigues,~Fred~S.~R.~Pinheiro, ~Gutembergue~S.~da~Silva,~Halysson~B.~Mendonça,~Fernanda~E.~S.~Galdino~and~Ricardo~Q.~de~F.~H.~Silva 
\thanks{Halysson Mendonça (orcid: 0000-0002-9411-2809) is with National Telecommunication Agency (ANATEL), Brazil (e-mail: hallyson@anatel.gov.br). 
Júlia da L. A. Silva (orcid: 0000-0003-0980-039X), Vicente A. de Sousa Jr. (orcid: 0000-0003-2859-6136), Marcio E. C. Rodrigues (orcid: 0000-0002-8399-3241), Fred S. R. Pinheiro (orcid: 0000-0002-4442-8784), Gutembergue S. da Silva (orcid: 0000-0001-5951-4805), Fernanda E. S. Galdino (orcid: 0000-0002-4462-6676) and Ricardo Q. de F. H. Silva (orcid: 0000-0003-0861-4341) are with Federal University of Rio Grande do Norte, Brazil (e-mails: 
    \{julia.andrade.097, vicente.sousa, marcio.rodrigues, fred.rossiter, gutembergue.soares, ricardo.queiroz.105, fernanda.galdino.075
\}@ufrn.edu.br).} 
\thanks{Submission: 2024-04-26, First decision: 2023-07-02, Acceptance: 2024-09-23, Publication: 2024-09-27.}
\thanks{This study was financed \ricrev{in part by} Coordena\c{c}\~{a}o de Aperfei\c{c}oamento de Pessoal de N\'{i}vel Superior - Brasil (CAPES) - Finance Code 001.} 
\thanks{Digital Object Identifier: 10.14209/jcis.2024.16.}
}

\maketitle


\begin{acronym}
  \acro{1G}{First Generation}
  \acro{2G}{Second Generation}
  \acro{3G}{Third Generation}
  \acro{4G}{Fourth Generation}
  \acro{5G}{Fifth Generation}
  \acro{5GRA}{Remote Areas applications}
  \acro{5GPHY}{5G Physical Layer}
  \acro{ADC}{analogic-to-digital converter}
  \acro{AGC}{automatic gain control}
  \acro{ASIP}{Application Specific Integrated Processors}
  \acro{AWGN}{additive white Gaussian noise}
  \acro{BDTM}{burst data transfer mode}
  \acro{BER}{bit error rate}
  \acro{BS}{base station}
  \acro{CDTM}{continuous data transfer mode}
  \acro{CFO}{Carrier Frequency Offset}
  \acro{CHF}{Characteristic Function}  
  \acro{CoMP} {Cooperative Multi-point}
  \acro{CP}{cyclic prefix}
  \acro{CR}{Cognitive Radio}
  \acro{CS}{cyclic suffix}
  \acro{CSI}{channel state information}
  \acro{CSMA}{carrier sense multiple access}
  \acro{DFT}{discrete Fourier transform}
  \acro{DPD}{digital pre-distortion}
  \acro{DZT}{discrete Zak transform}
  \acro{eMBB}{enhanced mobile broadband}
  \acro{EPC}{evolved packet core}
  \acro{FBMC}{Filter-bank multi-carrier}
  \acro{FDE}{frequency-domain equalizer}
  \acro{FDMA}{frequency division multiple access}
  \acro{FD-OQAM-GFDM}{frequency-domain OQAM-GFDM}
  \acro{FEC}{forward error control}
  \acro{FPGA}{Field Programmable Gate Array}
  \acro{FTN}{Faster than Nyquist}
  \acro{FT}{Fourier transform}
  \acro{FSC}{frequency-selective channel}
  \acro{GFDM}{Generalized Frequency Division Multiplexing}
  \acro{GS-GFDM}{guard-symbol GFDM}
  \acro{HPA}{high power amplifier}
  \acro{IBI}{inter-block interference}  
  \acro{ICI}{inter-carrier interference}
  \acro{IDFT}{Inverse Discrete Fourier Transform}
  \acro{IFI}{inter-frame interference}
  \acro{IMS}{IP multimedia subsystem}
  \acro{IoT}{Internet of Things}
  \acro{IP}{Internet Protocol}
  \acro{IQ}{in-phase and quadrature}
  \acro{ISI}{inter-symbol interference}
  \acro{IUI}{inter-user interference}
  \acro{KPI}{key performance indicator}
  \acro{LDPC}{low density check parity code}
  \acro{LLR}{log-likelihood ratio}
  \acro{LMMSE}{linear minimum mean square error}
  \acro{LTE}{Long-Term Evolution}
  \acro{LTE-A}{Long-Term Evolution - Advanced}
  \acro{M2M}{Machine-to-Machine}
  \acro{MA}{multiple access}
  \acro{MAC}{medium access control layer}
  \acro{MF}{Matched filter}
  \acro{MIMO}{multiple-input multiple-output}
  \acro{MMSE}{minimum mean square error}
  \acro{MRC}{maximum ratio combiner}
  \acro{MSE}{mean-squared error}
  \acro{mMTC}{massive machine type communication}
  \acro{MTC}{machine type communication}
  \acro{MU}{multi user}
  \acro{NEF}{noise enhancement factor}
  \acro{NFV}{network functions virtualization}
  \acro{OFDM}{Orthogonal Frequency Division Multiplexing}
  \acro{OOB}{out-of-band}
  \acro{OQAM}{Offset Quadrature Amplitude Modulation}
  \acro{PAPR}{peak to average power ratio}
  \acro{PHY}{physical layer}
  \acro{PRBS}{Pseudo Random Bit Sequence}
  \acro{PSD}{Power Spectrum Density}
  \acro{QAM}{quadrature amplitude modulation}
  \acro{QPSK}{quadrature phase shift keying}
  \acro{QoE}{Quality of Experience}
  \acro{QoS}{Quality of Service}
  \acro{RC}{raised-cosine}
  \acro{RF}{radio frequency}
  \acro{ROF}{roll-off factor}
  \acro{RRC}{root raised cosine}
  \acro{SC}{single carrier}
  \acro{SC-FDE}{Single Carrier Frequency Domain Equalization}
  \acro{SC-FDMA}{Single Carrier Frequency Domain Multiple Access}
  \acro{SCD}{Successive Cancellation Decoding}
  \acro{SDN}{software-defined network}
  \acro{SDR}{software-defined radio}
  \acro{SDW}{software-defined waveform}
  \acro{SEP}{symbol error probability}
  \acro{SER}{symbol error rate}
  \acro{SIC}{successive interference cancellation}
  \acro{SISO}{single-input single-output}
  \acro{SMS}{Short Message Service}
  \acro{SNR}{signal-to-noise ratio}
  \acro{ST}{space-time}
  \acro{STO}{Symbol Timing Offset}
  \acro{STC}{space time code}
  \acro{STFT}{short-time Fourier transform}
  \acro{TD-OQAM-GFDM}{time-domain OQAM-GFDM}
  \acro{TR-STC}{time-reversal space-time coding}
  \acro{TR-STC-GFDMA}{TR-STC Generalized Frequency Division Multiple Access}
  \acro{TVC}{time-variant channel}
  \acro{TVWS}{TV white space}
  \acro{UHF}{Ultra High Frequency}
  \acro{URLL}{ultra-reliable low latency}
  \acro{V2V}{vehicle-to-vehicle}
  \acro{VHF}{Very High Frequency}
  \acro{V-OFDM}{Vector OFDM}
  \acro{ZF}{zero-forcing}
  \acro{W-GFDM}{windowed GFDM}
  \acro{WHT}{Walsh-Hadamard Transform}
  \acro{WLAN}{wireless Local Area Network}
  \acro{WLE}{widely linear equalizer}
  \acro{WLP}{wide linear processing}
  \acro{WRAN}{Wireless Regional Area Network}
  \acro{WSN}{wireless sensor networks}
  \acro{TLS}{transport layer security}
  \acro{BRICS}{Brazil, Russia, India, and South Africa}
  \acro{LPWAN}{low-power wide-area network}
  \acro{UE}{user equipment}
  \acro{GPS}{Global Positioning System}
  \acro{RFID}{Radio-frequency identification}
  \acro{CAN1}{Controller Area Network 1}
  \acro{CAN2}{Controller Area Network 2}
  \acro{P2P}{Peer-to-peer}
  \acro{ICIC}{inter-cell interference coordination}
  \acro{ACK}{acknowledgment}
\end{acronym}

\begin{abstract}
Non-ionizing radiation is the subject of several applications, such as the microwave oven, which is very popular. Its use is characterized by direct user operation and reduced distances between users and the NIR source. Therefore, the level of human exposure in this scenario deserves attention as it is a potential cause of biological effects. In this sense, this paper presents the measurement of a microwave oven with a high level of NIR (above the limit defined by the standard) before and after two types of suggested repairs: the use of epoxy putty (Repair 1) and the replacement of the external surface of the oven (Repair 2). The definitions of measurement methodology, based on FDA and INMETRO standards, are presented and discussed. Finally, we describe an analysis of the efficiency of these repair methods \ricrev{aiming at safety operation}. 
\end{abstract}


\section{Introduction}

With technological advances, the use of Non-Ionizing Radiation (NIR) has gained prominence in several areas and applications. For this reason, there is concern about human exposure in an environment composed of different sources of NIR. Therefore, it is crucial to monitor and control exposure levels based on current standards.

The use of microwave ovens deserves to be highlighted because users directly operate the equipment, in addition to being at reduced distances from the NIR source. At the same time, the effects of exposure to microwave ovens are already the subject of studies, suggesting that it is the cause of biological variations~\cite{dondoladze2020}.

\ricrev{In~\cite{JEWA2024} (to be published)}, we show several NIR measurements from microwave ovens in residential usage, considering their location, brand, age, and state of conservation. From several measurement rounds, we have not found NIR value above the recommendations. Similar findings come from several other works~\cite{Zeyad_2001, Panait_2019}. 

However, similar to~\cite{Adnan_2013} and~\cite{Muhammad_2011}, a recent measurement caught our attention because it presented a high NIR level (\ricrev{above the standard indication}). For this reason, this work aims to present and discuss such measurements before and \ricrev{after repair, and also analyze the efficiency of different repair methods.}

 
This paper is organized as follows. Section~\ref{sec:nir} briefly presents the concept of NIR and current legislation. \ricrev{Section~\ref{sec:anatomy} describes the architecture and operation} of the microwave oven. \ricrev{Section~\ref{sec:measurement}} describes the measurements performed and presents the results and discussions. Final considerations are presented in Section~\ref{sec:conclusion}.

\section{Non-ionizing radiation and legislation}
\label{sec:nir}

The concept of radiation is associated with the natural and physical phenomenon that involves the transmission of energy through the propagation of electromagnetic waves through space or physical medium. Radiation is also divided into two groups: ionizing and non-ionizing. \ricrev{The difference between these groups is the frequency or wavelength they have in the electromagnetic spectrum} and, consequently, their effects on matter.

Non-ionizing radiation, the focus of this work, is defined as electromagnetic waves whose electric and magnetic fields have enough energy to raise only the excited state of electrons without ionizing matter. It is characterized by having an energy per photon of less than approximately 12 eV and wavelengths greater than 100 nm (frequencies less than $3 \cdot 10^{15} $ Hz)~\cite{ICNIRP-1998}. Therefore, NIR covers the range of low frequencies, radio waves, microwaves, infrared, visible light, and low-energy ultraviolet (UV-A) and plays essential roles in different areas of science and technology.

\ricrev{At a global level, ICNIRP is the main} international organization that sets regulations on radiofrequency exposure. \ricrev{The first organization guidelines dates back to 1998~\cite{ICNIRP-1998}; a revision was made in 2020~\cite{ICNIRP-2020}}. The legislation governing radiation levels relating to microwave ovens is a little more specific. Many countries adopt the approaches defined by the \textit{Food and Drug Administration} (FDA)\footnote{Federal Agency of the United States Department of Health and Human Services.}, such as Brazil, through INMETRO (National Institute of Metrology, Quality, and Technology)~\cite{INMETRO}. The limit is adopted as a radiation leakage of 50~W/m$^2$ at a distance of 5~cm~\cite{FDA1992}. \ricrev{FDA also outlines how specifications should be tested in microwave ovens}, which will be discussed in more detail in Section~\ref{sec:measurement}.

\section{\ricrev{BASIC ARCHITECTURE OF MICROWAVE OVENS}}
\label{sec:anatomy}

\begin{figure}[h]
     \centering
     \includegraphics[width=\linewidth]{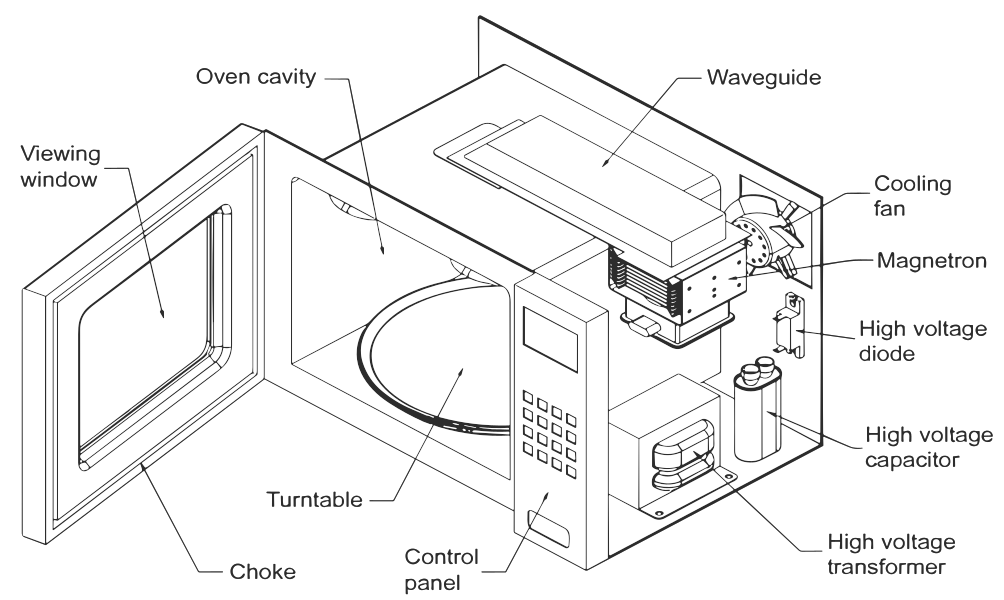}
      \caption{Illustration of the \ricrev{architecture} of \ricrev{a} microwave oven~\cite{oven-image}.}
     \label{fig:microwave-anatomy}
\end{figure}

\ricrev{Microwave ovens operate in} the microwave frequency range, as the name implies. Following international standards, the equipment operates at a frequency of 2.45 GHz with high power (\ricrev{ranging from about 600 to 1200 watts}). The components of the microwave oven are illustrated in Fig.~\ref{fig:microwave-anatomy}. They are generally divided between two functions: helping to cook food and ensuring that \ricrev{radiation does not leak out of the oven}.

\ricrev{The heat generated for cooking comes from intermolecular friction force}. Polar molecules, such as water, present in food are excited by an alternating electric field, \ricrev{moving quickly, so that the friction between them produces heat~\cite{vollmer2004}}. \ricrev{The magnetron is responsible for generating microwave energy that is transmitted to the cooking cavity through the waveguide}.

To prevent radiation leakage, the heating cavity is a thick-walled stainless steel structure that works like a Faraday cage, \ricrev{blocking microwaves from leaking out of the oven}. Furthermore, the equipment door is made, in addition to the glass, of a frequency selective surface, which is designed to act by \ricrev{blocking the passage of microwaves but allowing the passage of shorter wavelengths}. For this reason, there is the passage of light (\ricrev{which has shorter wavelength than microwaves}). Another essential safety function is that opening the door triggers a mechanism \ricrev{to cut off the power supplied to the magnetron}.

\section{Measurement}
\label{sec:measurement}

In Brazil, quality regulations and limit configurations for microwave ovens are established by ordinance number 174 of April 10, 2012~\cite{INMETRO} \ricrev{of INMETRO, which agrees to what is established in~\cite{FDA1992} by FDA}. Thus, based on the procedure defined by INMETRO, the measurement methodology of this work establishes:

\begin{itemize}
\item Precision measurements must be made with the microwave oven operating at its maximum power;
\item A load of 275 +/-15 milliliters of tap water must be used initially at 20~+/-2~degrees centigrade deposited in a borosilicate glass container with an internal diameter of approximately 85 mm, placed in the center of the load holder (the surface provided by the oven manufacturer);
\item The measurement must be made with an instrument that reaches 90\% of its steady state reading in 2 or 3 seconds when indicated at the intensity of the leak signal;
\item The equivalent power density existing in the vicinity of the outer surface of the oven must not exceed 50 watts per square meter (50 W/m²) \ricrev{at any point 5 centimeters or more away from the outer surface of the oven}.
\end{itemize}

The measurement setup, fulfilling the Brazilian regulation documents, consists of the following equipment:

\begin{itemize}
      \item Narda NBM-520 Broadband Field Meter (100 kHz - 60 GHz)~\cite{narda};
      \item NBM E-Field-Probe EF 0691 (100 kHz - 6 GHz)~\cite{sonda};
      \item Borosilicate glass container with 85~mm diameter and 400~ml total capacity.
\end{itemize}

\begin{figure}[h]
        \centering
        \includegraphics[width=0.8\columnwidth]{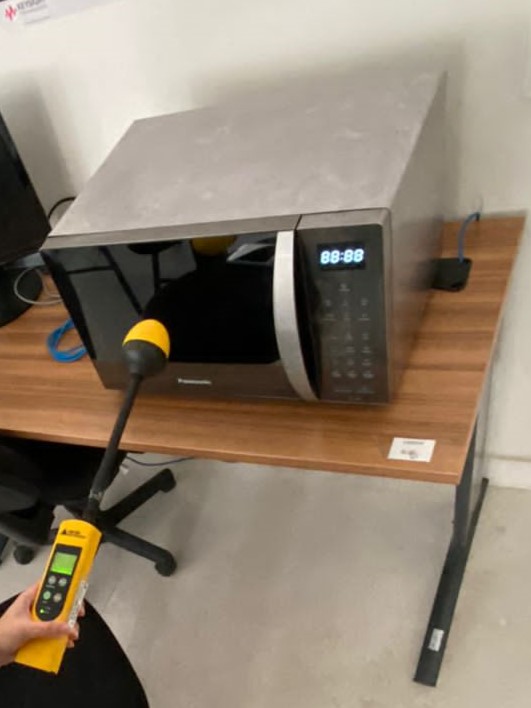}
        \caption{Measurement being carried out.}
        \label{fig:setup}
\end{figure}

As mentioned, this paper presents measurements carried out in a microwave oven that exceeded the limit defined by ~\cite{INMETRO}. Fig.~\ref{fig:setup} \ricrev{shows the measurement situation, with the handheld NBM-520 kept at the distance} of 5~cm between the oven's surface and the probe. The measurement procedure was as follows:

\begin{itemize}
    \item Carry out quick 10-second measurements on the different faces of the oven to identify where the highest emission is observed; 
    \item Take a 1-minute measurement in the highest-emission point; 
    \item Repeat the measurement with the oven turned off.
\end{itemize}

For this specific microwave oven, the highest emission region was \ricrev{in its front face}.

\ricrev{Turning off the oven, almost zero-power density levels were measured, showing that there was no contribution from external sources to the previous observed readings}.

\begin{figure}[h]
        \centering
        \includegraphics[width=0.8\columnwidth]{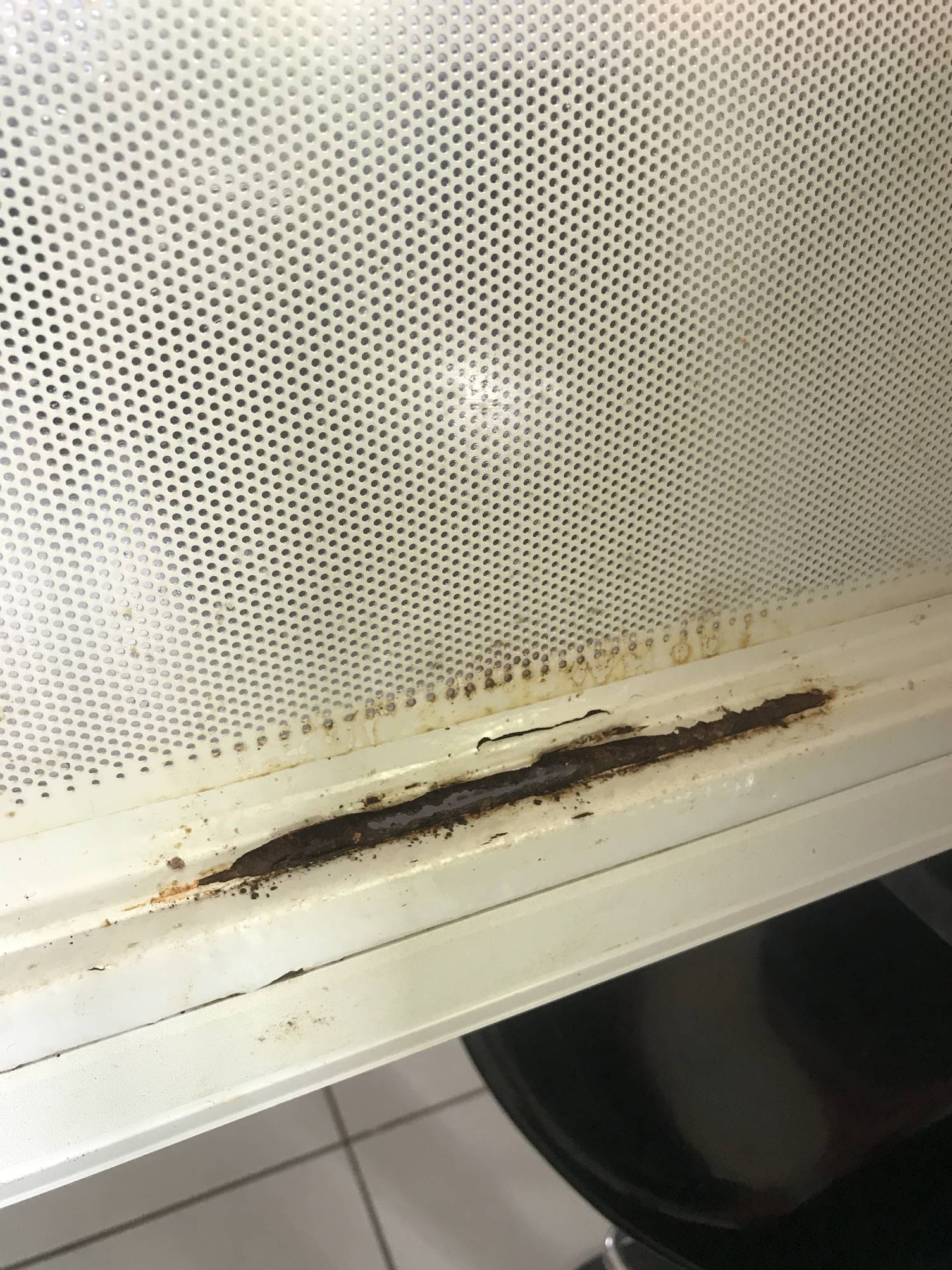}
        \caption{Initial state of the microwave oven door.}
        \label{fig:outlier-1}
\end{figure}

This equipment showed several signs of use, mainly rust, responsible for damaging the door structure, as shown in Fig.~\ref{fig:outlier-1}. \ricrev{Possibly, this is the reason that led to the leakage, as long as higher power density} levels were measured in front of the door. To overcome this leak, two types of repairs were carried out and evaluated:

\begin{itemize}
      \item Repair 1 (in-house repair): use of epoxy putty to fill the opening in the door made by rust (repair carried out by the oven's owners without the prior knowledge of the researchers involved in this work);
      \item Repair 2: replacement of the entire external surface of the microwave oven (repair carried out in a specialized location).
\end{itemize}

\begin{figure}[h]
        \centering
        \includegraphics[width=0.8\columnwidth]{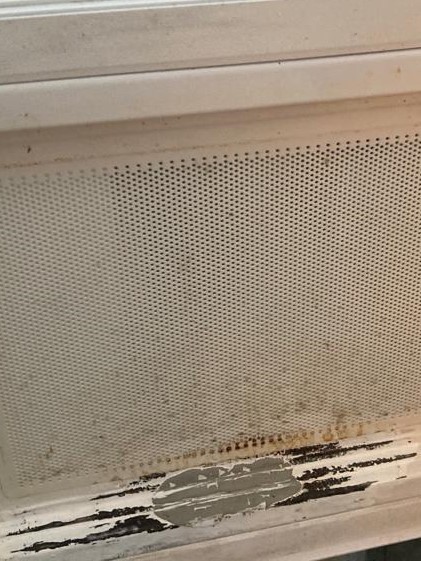}
        \caption{Microwave oven door internally: after Repair 1.}
        \label{fig:outlier-2}
\end{figure}

\begin{figure}[h]
        \centering
        \includegraphics[width=0.8\columnwidth]{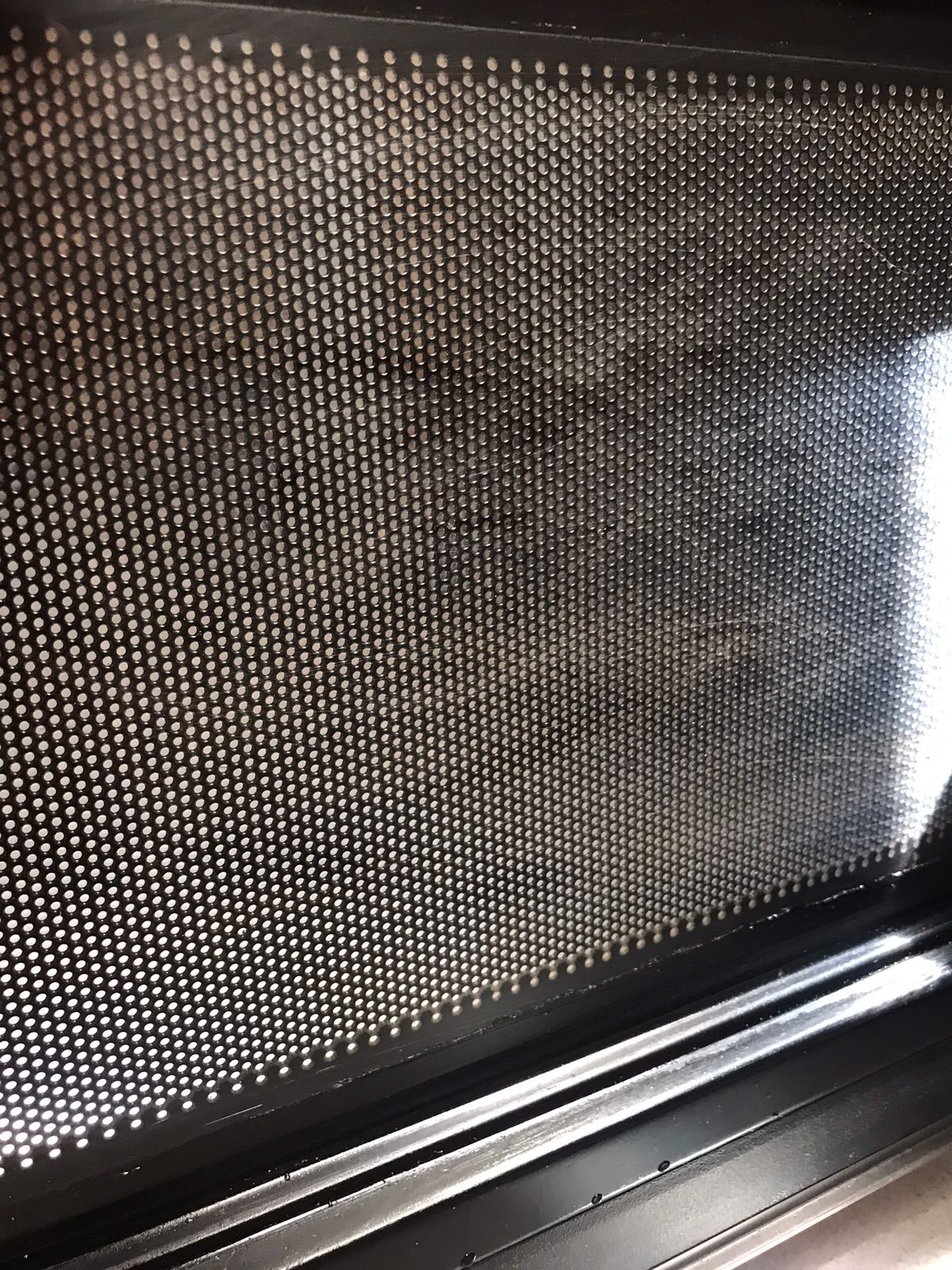}
        \caption{Microwave oven door internally: after Repair 2.}
        \label{fig:outlier-3}
\end{figure}

The measurements were divided into: initial state (Fig.~\ref{fig:outlier-1}), after Repair 1 (Fig.~\ref{fig:outlier-2}) and after Repair 2 (Fig.~\ref{fig:outlier-3}). The measured values of average and peak power density in each of them are described in Tab.~\ref{tab:powerdensity-micro-outlier}. We also show the exposure ratio (ER), the ratio between the measured value and the limit of 50 W/m$^2$, defined by regulation. 


\begin{table}[h]
    \centering
    \footnotesize
    \caption{Average and peak power density measured value.}
    \label{tab:powerdensity-micro-outlier}
	\begin{tabular}{cccc}
	\hline 
	\textbf{\makecell{State}} & 
    \textbf{\makecell{Average Power \\ Density W/m$^2$}} &
    \textbf{\makecell{Peak Power \\ Density  W/$^2$}} &
    \textbf{\makecell{ER (\%)}} 
    \\ \hline
    Initial & 183.00 & 416.30 & 832.60 \\
    After repair 1 & \footnotesize 37.50 & 111.50 & 223 \\
    After repair 2 & 0.18 & 0.42 & 0.84\\
    \hline 
	\end{tabular}
\end{table}

\ricrev{As reported in~\cite{JEWA2024}, we carried out measurements in a microwave oven of the same brand, model, year of acquisition, and with no visible signs of use.  The average power density corresponds to 0.12\% of the microwave oven outlier after the repair 2. Thus,} the NIR \ricrev{leakage} is more associated with the condition of the equipment than with the year of acquisition.


\section{Results Discussion and Conclusion}
\label{sec:conclusion}

This paper focuses on an outlier microwave oven with a measured power density value above 50~W/m² (the limit defined by national and international standards). Regarding the initial state of the microwave oven, the measured average and peak power density of 183~W/m² and 416~ W/m² represent 366\% and 832.6\% of the limit, respectively. Observing the state of conservation of the microwave oven, \ricrev{the presence of rust in the door structure is the main hypothesis} for high-emission results. \ricrev{We test our hypothesis with two types of repairs}: Repair 1, using epoxy putty to fill the door opening; and Repair 2, replacing the external surface of the oven.


When measuring again in front of the door after each repair, it was observed that, after Repair 1, the average power density, despite having decreased significantly, still assumes values very close to the limit and outside the reality of an equivalent microwave oven (same brand and year of acquisition), as exposed in~\cite{JEWA2024}. However, after Repair 2, exposure levels finally reached values within acceptable limits with an average of 0.18~W/m² and a peak of 0.42~W/m². Therefore, the importance of professional repair is reinforced to ensure the safe use of this device.

Finally, an important recommendation from the results of this paper is to take care of internal aspects of the microwave oven, especially \ricrev{related to rust in its} metal parts. This is especially important in coastal cities like Natal, \ricrev{where the measurements were carried out}.


\bibliographystyle{IEEEtran}
\bibliography{referencias.bib}
\end{document}